\documentclass[a4paper,10pt]{article}

\usepackage{colortbl}

\date{December 14th, 2023}

\usepackage{lineno}
\usepackage{xcolor}
\usepackage[utf8]{inputenc}
\usepackage{textalpha}
\usepackage{amsmath}
\usepackage{acronym}

\usepackage{hyperref}
\hypersetup{
hidelinks
}

\usepackage{tikz}

\usepackage[margin={2.5cm}]{geometry}

\usepackage{blindtext}

\usepackage{tabularx}
\usepackage{siunitx}
\newcommand{\myfig}[2]{\hyperref[#1]{\bfseries Figure~\ref*{#1}#2}}
\newcommand{\mysfig}[2]{\hyperref[#1]{\bfseries Supplementary Figure~\ref*{#1}#2}}

\usepackage{parskip}
\usepackage[superscript,biblabel]{cite}
\usepackage{doi}

\usepackage{caption}

\usepackage[]{graphicx}

\usepackage{lineno}
\usepackage{xspace}

\newcommand{\myref}[2]{\hyperref[#1]{\bfseries Figure~\ref*{#1}#2}}
\newcommand{\mysupp}[2]{\hyperref[#1]{\bfseries Supplementary Figure~\ref*{#1}#2}}

\begin{document}

\section*{A parallelized cellular Potts model that  
enables simulations at tissue scale}

Shabaz Sultan\textsuperscript{1,2}, 
Sapna Devi\textsuperscript{3},
Scott N. Mueller\textsuperscript{3},
Johannes Textor\textsuperscript{1,2}

\bigskip

\textsuperscript{1}  Data Science group,
Institute for Computing and Information Sciences,
Radboud University,
Nijmegen, The Netherlands 

\textsuperscript{2}  Department of Medical BioSciences, 
Radboud University Medical Center, 
Nijmegen, The Netherlands

\textsuperscript{3} Department of Microbiology and Immunology, 
The University of Melbourne, at the Peter Doherty Institute
for Infection and Immunity,  Melbourne, Victoria, Australia

\section*{Summary}

The \ac{CPM} is a widely used simulation paradigm for systems 
of interacting cells that has been used to study scenarios ranging from plant development
to morphogenesis, tumor growth and cell migration.
Despite their wide use, \ac{CPM} simulations are considered too computationally 
intensive for \ac{3D} models at organ scale.
\acp{CPM} have been difficult to parallelise because of their 
inherently sequential update scheme. Here, we present a \ac{GPU}-based
parallelisation scheme that preserves local update statistics and is up 
to 3-4 orders of magnitude faster than serial
implementations. We show several examples where our scheme preserves simulation
behaviors that are drastically altered by existing parallelisation methods.
We use our framework to construct tissue-scale models of liver and lymph node environments 
containing millions of cells that are directly based on microscopy-imaged tissue structures.
Thus, our \ac{GPU}-based \ac{CPM} framework enables \emph{in silico} 
studies of multicellular systems of unprecedented scale.

\clearpage

\section*{Introduction}

The \ac{CPM} \cite{PhysRevLett.69.2013} is a powerful simulation method that is known for its ability to simulate a wide range of biological scenarios. When modelling tissue environments, many researchers wish to consider both the complex morphology of single cells, and the behavior of these cells as part of larger collectives. The appeal of the \ac{CPM} lies partly in its ability to integrate both these aspects in a single model. As such, it has been applied to diverse problems including tumor growth \cite{10.3389/fonc.2013.00087}, vascularization \cite{Merks2008}, wound healing \cite{SCIANNA201533}, and morphogenesis \cite{hirashima,Maree3879}. Many extensions and hybrid model variants are available in the literature to, for example, add support for cell motility \cite{actmodel, thuroff, beltman}, cell signalling \cite{10.1371/journal.pcbi.1006919}, and finite-element based mechanical interactions \cite{RENS2020101488}. Thus, the \ac{CPM} can be described as a generic platform upon which a wide range of models can be built, possibly explaining its popularity in the tissue modelling field. 

For computational reasons, \ac{CPM} models are often built using a small number of cells (perhaps in the thousands) in simplified \ac{2D} geometries. Some research questions in tissue modelling however demand a realistically large number of cells, and accurate \ac{3D} representations of tissues. For instance, it is widely appreciated that the efficiency of space exploration of motile cells depends critically on the dimensionality of the environment, leading researchers to use \ac{3D} \ac{CPM} simulations \cite{beltman, wolf, vroomans}. Assessment of tumor growth behavior and invasive patterns of tumors likewise changes as a  function of dimension, resulting in the use of \ac{3D} \ac{CPM} simulations in the tumor modelling field as well \cite{10.3389/fonc.2013.00087}. While there is thus a clear research demand for realistically large \ac{3D} \ac{CPM} simulations, scaling up such simulations is generally considered computationally infeasible \cite{VanLiedekerke2015, reviewCellComputationalModeling, individualBasedModelling}. Removing this computational barrier would allow a host of \ac{CPM} based modelling research to be scaled to what is biologically realistic, instead of scaling to what technical constraints allow.

The \ac{CPM} is an on-lattice method: simulations are executed on a grid. A common strategy to speed up grid-based models is to cut the simulation into smaller subsections that are run in parallel. However, for the \ac{CPM}, model behavior is intimately tied to locations on the grid being updated sequentially, in a single-location update Markov chain Monte Carlo process. As such, parallelisation has been cautioned against, as it risks drastic changes in \ac{CPM} ``system kinetics'' \cite{hirashima}. Nevertheless, several attempts have been made to employ subsection-based parallelisation. Chen \emph{et al.} proposed a checkerboard scheme that divides the \ac{CPM} lattice up over a computer cluster, with sections chosen much larger than cells, preventing cell-level synchronization issues \cite{CHEN2007670}. Tapia \emph{et al.} propose a similar scheme for \acp{GPU} \cite{TAPIA2011857}: because the sections are much smaller than a cell on the \ac{GPU},  they use a per-cell locking mechanism to prevent simultaneous updates of a cell.  Neither approach has been extensively checked for correct model behavior. To our knowledge, no publicly available implementation of subsection-based parallelisation exists for the \ac{CPM}. The software ``Morpheus'' \cite{morpheus} includes a non-subsection-based \ac{CPU} parallelisation strategy; this method is less suited to fully utilise parallel accelerator hardware such as \ac{GPU}s. 

In this paper, we aim to make several contributions: First, we introduce a highly optimized parallelized implementation of the \ac{CPM} using several optimization strategies not previously explored in the literature. Second, we conduct an empirical analysis of potential pitfalls in parallelising a model dependent on serial updates of single locations and show that our approach mitigates these issues while preserving performance benefits. Third, we demonstrate that our approach retains the behavior of cell motility simulations that are particularly prone to changed model kinetics upon parallelisation. And finally, we build two large \ac{3D} image-based simulations of T~cell migration in liver and lymph nodes, demonstrating how complex tissue environments can be analysed at a larger scale and realism than was possible before.

\section*{Results}

\subsection*{Serial Cellular Potts Models are limited to sub-millimetre scale}

\begin{figure}[]
\centering
\includegraphics{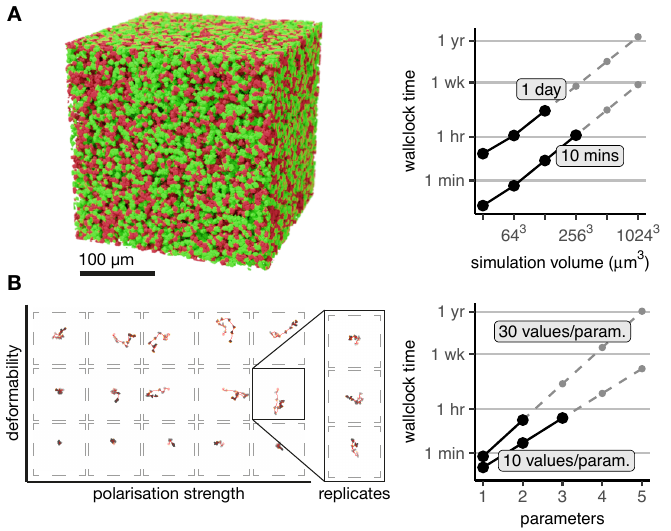}
\caption{{\bfseries Serial \ac{CPM} implementations do not scale well to large simulations and big ensembles.} 
    {\bfseries (A)} Computational cost of simulating a 256\textsuperscript{3}$\mu$m\textsuperscript{3} volume (1$\mu$m\textsuperscript{3} per voxel) for 10 simulated minutes (1s per simulation step). With a high-performance CPU implementation, this takes  1.2 hours of wall-clock time. Extrapolation to a full day of simulated time for a 1mm\textsuperscript{3} simulation estimates 1.4 years of wall-clock time. Time measurements (black) are averages of 5 simulations, projections are based on linear fits of simulations $\geq$ 64\textsuperscript{3}$\mu$m\textsuperscript{3} in log-log space (grey). 
    {\bfseries (B)} Building \acp{CPM} typically requires tuning parameters. Here we perform a multidimensional grid search for parameters that yield motile cells using the Act extension of the \ac{CPM} \cite{actmodel}; a modest single parameter grid search where 10 values are tried takes 15.9 seconds; larger grid searches take exponentially more time as function of parameters searched, with a five parameter search with 30 values is extrapolated to 1.1 years of wall-clock time. Time measurements (black) are averaged over 5 simulations, projections fitted on linear trends in linear-log space. \ac{CPM} parameters are shown in Table~\ref{table_model_settings_2d} and Table~\ref{table_model_settings_3d}.}
\label{cpuscaling}
\end{figure}

The \ac{CPM}'s serial update schedule -- only one location at a time is updated on a grid -- is central to its behavior, and thus not trivially changed. We therefore first aimed to quantify the computational limitations for models based on the serial \ac{CPM}: how big can our computational experiments be, and what timescale can they span, without making fundamental changes to the \ac{CPM}? 

We set up \ac{3D} \ac{CPM} tissue experiments at different spatial and temporal scales and measured the required wall-clock time. We choose spatial and temporal simulation resolutions that allow reasonably accurate representations of highly motile cells within tissues. Lymphocytes are among the smallest nucleated cells in the mammalian body at 8-10μm diameter for B~cells and T~cells. At speeds of up to a few body lengths per minute, they are also among the fastest movers within tissues. We chose 1μm\textsuperscript{3} per voxel and 1 second per simulation step, or \ac{MCS}, as our simulation resolutions, corresponding to Beltman \emph{et al.} \cite{beltman}.

To create realistically large models of, for example, a complete murine \ac{LN}, or scale a tumor model to the size of an avascular tumor spheroid \cite{shirinifard}, experiments at mm\textsuperscript{3} scale would be needed. Certain organ scale dynamics require timescales of at least a day, such as T~cell transit through \acp{LN} \cite{Mandl18036}. More modest simulation experiments at 256\textsuperscript{3}μm\textsuperscript{3} (\myfig{cpuscaling}{A}) at 10 simulated minutes require \textasciitilde 1h of wall-clock time on a highly optimized \ac{CPU} implementation of the \ac{CPM} (Methods). Such experiments are sufficient to see minimal movement of motile cells such as T~cells and can capture organ compartments such as a \ac{LN} paracortical T~cell zone, or a micrometastasis for cancer modelling. For  mm\textsuperscript{3} scale, one-day experiments however, researchers would have to wait at least a year -- making them impractical for all but the most patient of researchers. 

A single experiment is rarely sufficient, in both empirical and simulation research. Beyond the need for experimental replicates, simulations can have parameters for which associated model behavior is non-obvious, requiring parameterized simulation ensembles. In \myfig{cpuscaling}{B}, we explore a cell motility extension to the \ac{CPM} that coarsely models actin dynamics \cite{actmodel}. A single two-dimension simulation is relatively cheap, but when exploring even a modest number of parameters the execution time goes up exponentially. 

Thus, serial \ac{3D} \ac{CPM} simulations at a scale of 1~mm\textsuperscript{3} are currently not feasible, and even 100\textsuperscript{3}μm\textsuperscript{3} scale simulations were extremely challenging in our experiments. While ensemble runs can be accelerated with parallel simulations without changing the \ac{CPM}, the high computational costs of running large ensembles further motivates the need for a faster CPM.

\subsection*{Designing a parallel execution scheme for the \ac{CPM}}

To make large scale simulations practical, we propose a parallelisation scheme for the \ac{CPM} that uses highly parallel \ac{GPU} hardware. A natural way to parallelise on-lattice models, such as the \ac{CPM}, is to divide the simulation grid in smaller sections; each section runs a local simulation in parallel. As \ac{GPU}s support up to thousands of calculations in parallel, these sections will be quite small. The \ac{CPM} also updates information outside of the grid, such as sizes of simulated cells. We first consider three issues when designing our parallelisation scheme: adjacent sections with active simulation interfering with each other during updates of the simulation grid, non-grid information such as cell sizes being updated simultaneously, and stale information being used by one parallel simulation after said information got updated by another simulation.

\begin{figure}[]
\centering
\includegraphics{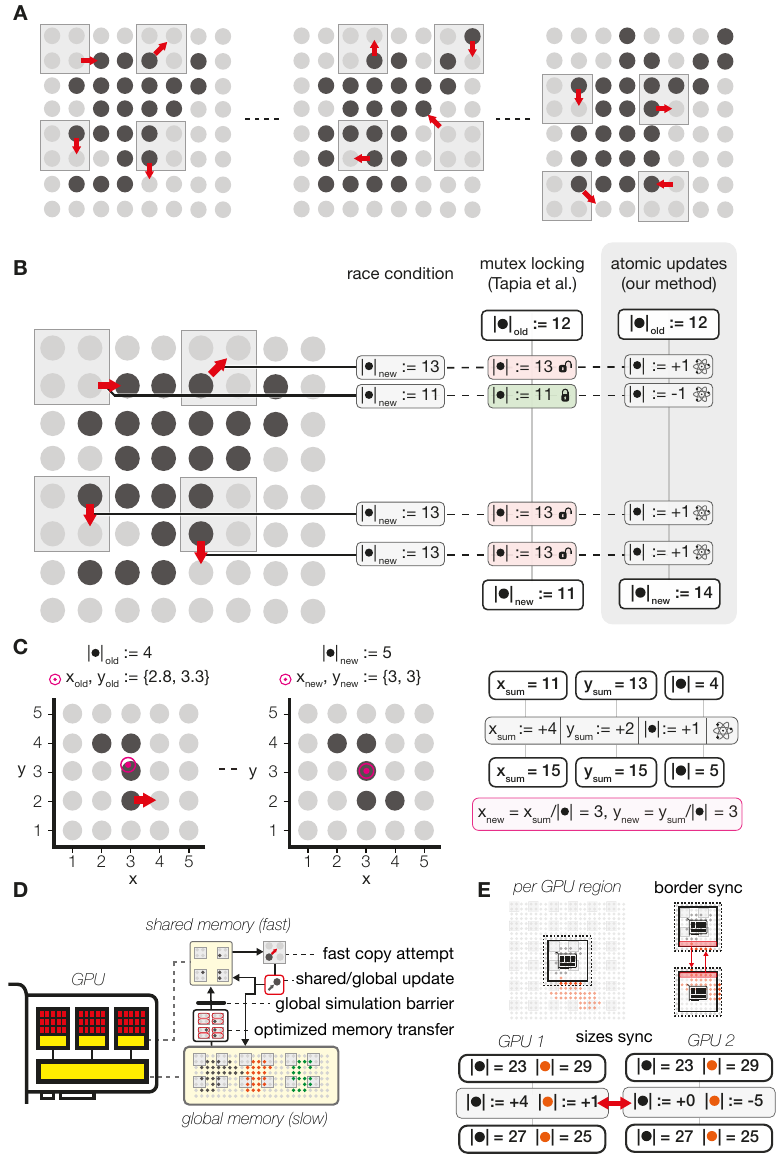}
    \caption{{\bfseries Designing a parallel \ac{CPM} based on checkerboarding and atomic updates for \acp{GPU}.}
    {\bfseries (A)} We divide simulation space into a subgrid, with alternating regions being active in a ``checkerboard'' pattern, preventing update conflicts on the grid. 
    {\bfseries (B)} We implement two strategies to prevent global variable update conflicts (e.g. cell size tracking): a mutex locking strategy (Tapia \emph{et al.} \cite{TAPIA2011857}), and a novel atomic operation strategy.
    {\bfseries (C)} Combining interleaved storage and atomic operations on adjacent data allows variables to be kept in sync with each other.
    {\bfseries (D)} A four-step execution scheme, where active checkerboard areas are transferred to fast shared memory, accelerates copy attempts. Thread barriers synchronize active threads after transfers.
    {\bfseries (E)} A multi-\ac{GPU} execution scheme, where regions of the simulation are assigned to different \acp{GPU}. Borders are synchronized between relevant adjacent regions, and global variables are synchronized through delta tracking.}
\label{parallel_scheme}
\end{figure}

A \ac{CPM} simulation proceeds by choosing a random position on the simulation grid, and stochastically copying the position’s state to a neighbour. If adjacent subsections of the grid each have an active parallel simulation running, we run the risk of simulations ``racing’’ to write to the same location on the sections’ boundary. Such ``race conditions’’ run the risk of information getting lost, as one section overwrites an almost simultaneous update – without incorporating information of the overwritten update. This is a standard problem in parallelisation, with a straight-forward solution: never allow two adjacent sections to be active at the same time. A spaced out ``checkerboard’’ pattern of sections is only ever active; active sections are sufficiently spaced to prevent overlapping updates (\myfig{parallel_scheme}{A}). Regions cycle periodically between active and dormant state. 

The CPM also tracks information involving more than one grid location, such as the current sizes of simulated cells. We face the same ``race condition’’ problem here, as it is possible that multiple parallel updates touch a single cell (\myfig{parallel_scheme}{B}, left column). \ac{GPU} parallelisation requires a high number of small active subsections, making such concurrent updates more likely.  In Tapia \emph{et al.}\cite{TAPIA2011857} a per-cell locking scheme is proposed, that rejects all but the first concurrent update attempt on a variable  (\myfig{parallel_scheme}{B}, middle column). We instead opt to use atomic operations (\myfig{parallel_scheme}{B}, right column), which equally prevents update information from getting lost, as the hardware will eventually integrate all updates for such operations.

In serial CPMs, global information travels instantaneously: as only one grid position is updated at a time, every update attempt has access to the results of all previous updates. This instantaneous information access is relaxed in our proposed scheme: when information such as cell size is accessed, this information may be out-of-date due to still in-flight updates to the accessed information. The cell-locking scheme from Tapia \emph{et al.} prevents this from happening, by outright rejecting any updates that could have used out-of-date information; we explored a less strict scheme, reasoning that the stochastic CPM copy attempts may tolerate out-of-date information. For CPM extensions that do not tolerate their global variables being slightly out sync with each other (e.g. variables used in center-of-mass ``centroid'' tracking) we designed an additional scheme that still will keep variables in sync with each other, and sits on top of our atomics based scheme (\myfig{parallel_scheme}{C}, Methods).  Otherwise, due to atomic updating, our scheme still gives strong guarantees on tracked data being correct – for example, cell sizes will be tracked correctly – and the added flexibility allows us to explore the effects of allowing the use of slightly outdated information.

Implementation of our scheme on real hardware requires consideration of several technical details. On GPUs, memory access almost always bottlenecks performance -- but the CPM's random access pattern prevents easy use of a \ac{GPU}'s caching hierarchy \cite{TAPIA2011857}. To maximize performance, we designed a data transfer scheme that carefully operates in lock-step with  checkerboard work splitting (\myfig{parallel_scheme}{D}, Methods): we cache local data needed by the regions that are becoming active, and overlap memory transfer and simulation calculations to fully utilise GPU hardware. To increase simulation volume, and gain further speed ups, we also prototyped a multi-GPU scheme (\myfig{parallel_scheme}{E}, Methods).

In short, our proposed parallel CPM scheme is tailored to make optimal use of GPUs: hardware specialized in running parallel code. It still gives strong guarantees on simulation state consistency -- for example, a global variable tracking a cell's size will always equal the number of pixels on the grid occupied by that cell. Yet, we do allow the use of slightly outdated values of such global variables in copy attempts to gain flexibility and further maximise performance gains.

\subsection*{Aligning the lattice update statistics of parallel CPMs to serial CPMs}

\begin{figure}[]
\centering
\includegraphics{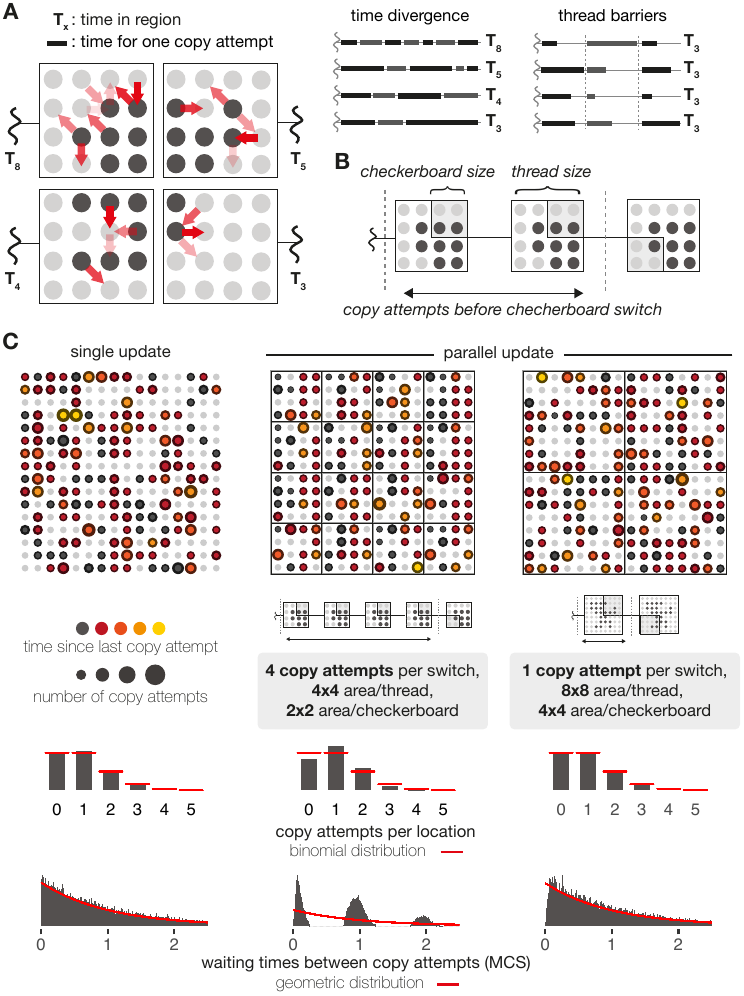}
    \caption{{\bf Checkerboarding can drastically alter local update statistics.}
      {\bfseries (A)} To prevent threads from racing ahead of others too much, thread barriers are imposed after each update. 
      {\bfseries (B)} Checkerboard size and switching frequencies determine the distribution of waiting times between consecutive updates of the same locations. 
      {\bfseries (C)} Small checkerboard areas or infrequent switches can alter the waiting time distribution substantially compared to the single-update scheme. Update frequency histograms were generated by simulating one MCS (256 updates) on a 16x16 grid. Waiting time histograms were simulated on the same grid using 100 simulated MCSs.
}
\label{update_statistics}
\end{figure}

Our novel parallel execution scheme guarantees simulation data consistency, but drastically changes the way the simulation lattice is updated. We investigated statistical properties of copy attempts on the simulation lattice for different lattice positions; specifically, we analysed the relative ordering of copy attempts, frequency of copy attempts, and waiting times between copy attempts. This allowed us to quantify the divergence of parallel lattice update kinetics from the serial CPM. Such divergence could lead to changed simulation behavior in complex models of cells and tissues.

Without guardrails, performing independent CPM updates in parallel would allow different parts of a simulated tissue to have been alive for different lengths of simulated time (\myfig{update_statistics}{A}). Modern GPUs use thread barriers to synchronize parallel threads without strong computational performance implications. We set thread barriers after ever single copy attempt, so for every currently active checkerboard region there is only a maximum time divergence of one copy attempt between active regions.

Our parallel execution scheme allows us to freely choose the size of subregions and frequency of active region switching (\myfig{update_statistics}{B}), without impacting data consistency guarantees. We measure frequency and waiting times between copy attempts per position on the lattice in the standard, serial implementation of the CPM (\myfig{update_statistics}{C}, left column). For our initial parameter choice, we found that underlying update statistics are drastically changed (\myfig{update_statistics}{C}, middle column): waiting times between copy attempts became much more predictable due to the much more even distribution of copy attempts compared to the serial CPM. Using large parallel subsections with frequent active region switching allows non-parallel update statistics to be closely recapitulated in our parallel scheme (\myfig{update_statistics}{C}, right column). 

We thus find that parallelisation parameters have major influence on copy attempt statistics on the lattice. By switching active regions of the  checkerboard frequently, and avoiding too small areas per 
thread, we can obtain statistics that closely resemble the serial CPM.

\subsection*{Emergent cell and tissue behavior is maintained in a parallel \ac{CPM}}

\begin{figure}[]
\centering
\includegraphics{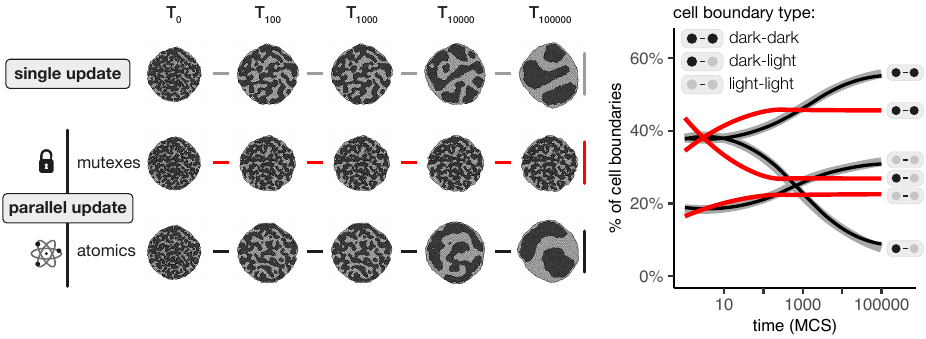}
    \caption{{\bf Atomic updates lead to correct behavior in a simulation of cell sorting.}
    A cell sorting simulation based on Graner \emph{et al.} \cite{PhysRevLett.69.2013} is implemented with two strategies to prevent race conditions for cell size variables. The mutex locking strategy proposed by Tapia \emph{et al.} \cite{TAPIA2011857} leads to severe starvation in our tests. Using an atomic update approach recovers the desired sorting behavior, closely matching sorting speed of the original \ac{CPM}. The simulation was run with a 4x4 area per thread and 2x2 checkerboards, but other sizes were tested and did not make a significant difference for this simulation. Sorting trends of different \ac{CPM} implementations are averaged over 5 simulations. Simulation parameters can be found in Table~\ref{table_model_settings_2d}.}
\label{sorting}
\end{figure}

\begin{figure}[]
\centering
\includegraphics{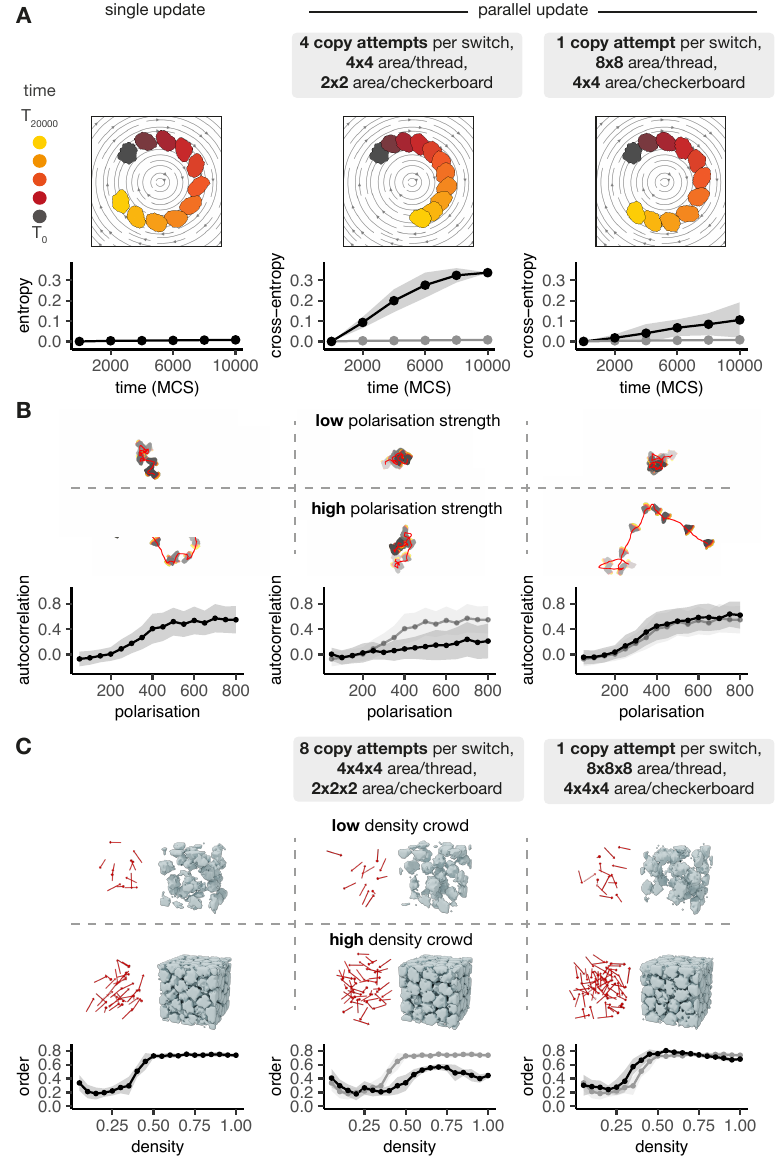}
    \caption{{\bfseries Motility behavior matches serial \ac{CPM} behavior when parallel \ac{CPM} local statistics are similar.}
    {\bfseries (A)} Cells in a circular chemokine gradient simulation move highly regular in a serial CPM (low per-pixel entropy, shaded: 1SD); parallel simulations diverge slowly from reference behavior when copy attempt statistics are preserved (low per-pixel cross-entropy).
    {\bfseries (B)} 
    In the Act model of intra-cell polarization \cite{actmodel}, stronger polarization strength leads to higher directional autocorrelation (measured for 750 \acp{MCS} at 50 \ac{MCS} intervals, averaged over 50 simulations, shading: $\pm$1SD); directional persistence only emerges when update waiting time distributions are preserved.
    {\bfseries (C)} Phase transition simulation in 3D where synchronous motility emerges at a density around 0.5. Cell order indices after 2000 \acp{MCS} are shown (averaged over 5 simulation runs, shading: $\pm$1SD). The phase transition only occurs when update waiting time statistics are preserved. Simulation parameters can be found in Table~\ref{table_model_settings_2d} and Table~\ref{table_model_settings_3d}.}
\label{fig6}
\label{motility_models}
\end{figure}

Our goal is to build parallel CPMs that are much faster than previous implementations but maintain cell and tissue kinetics consistent with existing serial CPMs. We hypothesized that aligning the waiting times between update attempts of our parallel CPM to the serial version should suffice to achieve such consistency. To investigate this, we ran several simulation scenarios with different CPM variants, measuring emergent behaviors on the cell and tissue levels and comparing these to serial CPM versions. 

We first investigated the classic cell sorting simulation \cite{PhysRevLett.69.2013}: two mixed populations of cells with differential adhesion (i.e., cells of the same type have stronger adhesion to each other than cells of different types) sort themselves in two homogeneous regions over time (\myfig{sorting}, grey lines). The mixing kinetics can be measured by tracking homogeneous and heterogeneous cell neighbours over time \cite{PhysRevLett.69.2013}. 
 Strikingly, the mutex based parallelisation approach proposed by Tapia \emph{et al.}\cite{TAPIA2011857} -- while still being able to run a simulation -- leads to a lack of sorting behavior; the simulation encounters so-called ``starvation’’, because only one concurrent copy attempt is allowed per cell under this scheme (\myfig{sorting}, red lines). Our less restrictive scheme allows expected sorting behavior to re-emerge (\myfig{sorting}, black lines). The sorting behavior remained consistent for all parallelisation parameters we examined (\mysupp{sorting_accuracy}). 

We next examined three motility model variants of the CPM, where additional biases are placed on copy attempts to obtain cells that move persistently; in the standard CPM movement only occurs as a function of random membrane fluctuations. Because motile cells have a leading and a retracting edge and need to roughly maintain their target size, copy attempts need to sufficiently alternate between these cell sections for movement to occur. We therefore speculated that movement models might be more affected by changes to the waiting times between copy attempts.

The first motility model we consider adds a chemotactic vector field to the CPM; copy attempts are more likely to succeed when aligned with the underlying vector field (see Methods). Because the directional bias is fixed in this model -- the field does not change over time -- cells move in a highly regular fashion, and simulations with the same initial conditions closely align  even at a pixel level (\myfig{motility_models}{A}, left column). We quantified this alignment using pixel-wise (cross-)entropy: we measure the probability that a pixel contains a cell in multiple simulations, where low entropy of pixel probabilities indicated high agreement between simulations, and low cross-entropy indicates a high level of agreement of parallel simulations with reference serial runs. In simulations with highly changed lattice update statistics, cross-entropy increased quickly (\myfig{motility_models}{A}, middle column); when the update statistics are preserved cross-entropy rises much slower (\myfig{motility_models}{A}, right column).

Next, we analysed a model that simulates localized cell polarization \cite{actmodel}, where recent pixel updates at the border of a cell makes further copy attempts involving those pixels more likely to succeed. The waiting times between copy attempts are very important in this model, as cells start to move when many pixels in a small neighbourhood successfully expand at the same time; this is less likely when copy attempts are spaced out evenly. A polarization strength parameter controls how much polarization bias is added to a cell; varying the polarization strength interpolates motility between Brownian and directionally persistent motility modes (\myfig{motility_models}{B}, left column). In parallel simulations with changed update frequency and waiting time statistics this full range of behaviors does not appear (\myfig{motility_models}{B}, middle column), but they re-emerge when update statistics are preserved (\myfig{motility_models}{B}, right column).

Finally, we examined a simulation of a tightly packed \ac{3D} collective of immune cells \cite{beltman}, where each cell has a preferred movement direction vector; copy attempts for a cell are more likely to succeed when they align with this vector, and the vector is periodically updated to align with recent movement of the cell. This motility model involves information tracked on the lattice, and additional information tracked external to the simulation grid, placing additional demands on a parallel CPM.  In a similar simulation model, low density cell collective exhibit Brownian movement, transitioning sharply to ordered collective movement as density is increased \cite{PhysRevE.74.061908}; this characteristic phase transition is reproduced in the CPM (\myfig{motility_models}{C}, left column). In a parallel CPM this phase transition disappears when update statistics are changed (\myfig{motility_models}{C}, middle column), but re-emerges when update statistics are preserved (\myfig{motility_models}{C}, right column).

Taken together, these results demonstrate that a strict update scheme preventing parallel updates of the same cell is not necessary for maintaining valid CPM behavior and can even lead to severe ``starvation'' where simulations do not progress anymore. On the other hand, preserving waiting times between copy attempts involving the same pixel can be very important to maintain model behavior, and was critical in all three motility scenarios we examined.

\subsection*{\ac{GPU}-based \ac{CPM} parallelisation 
	accelerates models by several orders of magnitude}

Having confirmed that our parallel \ac{CPM} indeed produced correct model behavior in a range of examples, we next investigated the performance improvements resulting from our implementation choices when compared to a serial \ac{CPM}. 

\begin{figure}[]
\centering
\includegraphics{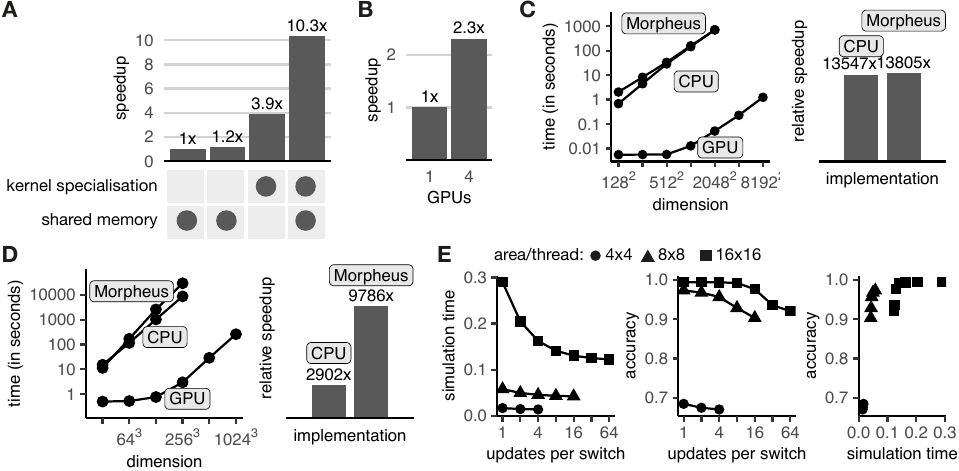}
\caption{{\bfseries Parallel CPMs outperform serial implementations by orders of magnitude.} 
    {\bfseries (A)} Performance benefits are obtained by a shared memory scheme, and a pre-compiled kernel specialization scheme, based on a sorting simulation extended to densely fill simulation space (\myfig{sorting}, all sorting benchmarks are average of 5 simulations of 100 MCSs, 100 MCS burn-in).
    {\bfseries (B)} Distributing the lattice over 4 \acp{GPU} speeds up the simulation further.
    {\bfseries (C)} Performance of sorting simulations are significantly higher relative to two serial implementations; relative speedup shown for 2048x2048 simulations.
    {\bfseries(D)} Performance of 3D cell collective simulations is also improved (\myfig{motility_models}{C}, average of 2 simulations of 100 \acp{MCS}, 100 \ac{MCS} burn-in); relative speedup shown for  256$^3$ simulations. 
    {\bfseries(E)} Trade-offs between speed and accuracy for chemokine field simulations (\myfig{motility_models}{A}, accuracy: 1 - cross-entropy of parallel ensemble relative to serial ensemble, 250 simulations per ensemble). CPM parameter settings can be found in Table~\ref{table_model_settings_2d} and Table~\ref{table_model_settings_3d}.}
\label{performance}
\end{figure}

We first examined the performance gains resulting from our use of a shared memory caching scheme (\myfig{parallel_scheme}{D}, Methods) using a 2D sorting simulation benchmark (\myfig{sorting}). On its own, shared memory caching achieved only minimal speedup (\myfig{performance}{A}). However, for further performance improvement, we pre-generated different versions of our code based on which \ac{CPM} extensions were in use (Methods). Combining these specialized kernels with the caching scheme achieved a significant 10-fold speedup, more than expected from either optimization alone (\myfig{performance}{A}). Testing our multi-\ac{GPU} scheme (\myfig{parallel_scheme}{E}, Methods) in the same scenario achieved a further 2.3x speedup by distributing the simulation across four \ac{GPU}s (\myfig{performance}{B}). Hence, careful consideration of \ac{GPU}-specific implementation choices resulted in significant performance benefits.

We next compared our parallel \ac{CPM} (on a single \ac{GPU}) to our own optimized \ac{CPU} implementation (Methods), and to the state-of-the-art software ``Morpheus'' \cite{morpheus}. Using the same 2D cell sorting benchmark, we observed a >3,500-fold speedup compared to our own CPU implementation, and a >25,000-fold improvement relative to Morpheus (\myfig{performance}{C}). On the 3D cell collective simulation (\myfig{motility_models}{C}), a >600-fold speedup was achieved relative to serial CPMs (\myfig{performance}{D}). It is important to note that our benchmark hardware did not permit use of shared memory caching for the 3D model; future hardware with more shared memory may improve 3D performance.
 
As shown in \myfig{update_statistics}{} and \myfig{motility_models}, checkerboard size and switching frequency influences model behavior: larger divisions and more frequent switching leads to behavior concordant with serial CPMs. These update patterns also interact with caching, further impacting performance. As the sorting simulation has low sensitivity to parallelisation parameters (\mysupp{sorting_accuracy}), we instead benchmarked our 2D chemotaxis simulation (\myfig{motility_models}{A}). We found that larger subdivisions indeed slowed simulations down considerably, whereas frequent switching led to a less marked slowdown (\myfig{performance}{E}). These results suggested a trade-off between performance and model accuracy. 

Hence, our parallel \ac{CPM} represents a significant speed improvement over even well optimized serial \ac{CPM}s such as Morpheus. This improvement was to a substantial extent a result of \ac{GPU}-specific optimizations and led to an overall speedup of several orders of magnitude.

\subsection*{Building an image-based model of T~cell surveillance of the liver}

Our parallel \ac{CPM} was motivated by the need for tissue simulations at realistic scale with realistic cell morphology. While our approach increases engineering complexity of the model, it can also simplify model building, as it enables direct use of imaging data to initialize the simulation environment. To determine the feasibility of directly using real-world imaging data in our system, we constructed a model of T~cells moving in the sinusoidal regions of the liver. 

\begin{figure}[]
\centering
\includegraphics{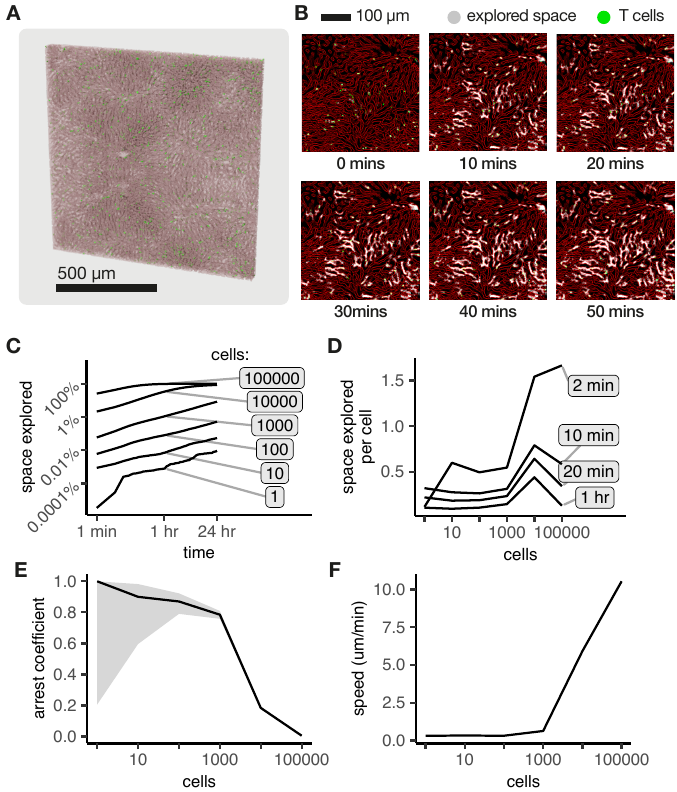}
      \caption{{\bf T cells in a liver sinusoid simulation exhibit a transition from a non-motile to a motile state.} 
      {\bfseries (A)} A 1024$\times$1024$\times$182 μm\textsuperscript{3} scan of a liver sinusoid is directly imported into our simulation environment (64μm slice shown). 
      {\bfseries (B)} Simulated T cells explore liver sinusoids (5000 cell simulation shown). 
      {\bfseries (C)} Simulations at different cell densities explore the environment at different speeds (single simulation of 24hr, with 1 min snapshots). 
      {\bfseries (D)} An increase of explored volume per cell occurs when cell density rises from 1,000 to 10,000 cells; 100,000 cells fill $\sim$1/3 of sinusoid volume, causing exploration efficiency to saturate.
      {\bfseries (E)} At low cell densities most simulated cells are in an un-polarized, non-motile state; at higher densities collisions cause cells to transition in to a polarized, motile state (1 simulation of 30 min per density, arrest coefficient: fraction cells < 2μm/min\cite{Hugues2004}, shaded: Wilson score interval based on the number of 2-minute subtracks). \ac{CPM} parameter settings can be found in Table~\ref{table_model_settings_liver_ln}.
      {\bfseries (F)} Average speed of the simulated T cells at different densities.}
\label{liver}
\end{figure}

Tissue resident memory T cells (Trms) are confined to specific organs, where they sample their environment for signs of disease \cite{masopust} -- in particular, liver Trms are key in providing sterilising immunity for malaria \cite{Lefebvre}. Trms display a distinct motility pattern in the liver, and efficacy of malaria clearing depends on the number of Trms present \cite{Fernandez-ruiz}. We constructed a \ac{3D} simulation environment of liver sinusoids -- capillaries that Trms reside in and migrate through -- and combined it with a polarized cell motility model \cite{actmodel} to represent the Trms (\myfig{liver}{B}). Trm motility parameters were derived from a \ac{LN} model constructed in parallel, and our parameter choice will be explained in more detail with the \ac{LN} data. These parameter settings lead to motility modes also observed in intravital liver data: intermittent movement and pausing of Trms. The goal of our simulation was to predict how much of the liver volume would be explored by a given number of T~cells within a set amount of time; this same question was addressed by Fernandez-Ruiz \emph{et al.} using a more abstract mathematical model \cite{Fernandez-ruiz}.

As can be expected, we found that liver exploration by the Trm population increased with the number of exploring cells, and with the amount of exploration time (\myfig{liver}{C}). Remarkably, the sinusoid volume explored by an individual cell increased dramatically as the total Trm population was raized from 1,000 to 10,000 cells (\myfig{liver}{D}). Further investigation revealed the source of this phase transition: cells in our simulations can be in an unpolarized state, in which they do not move persistently; collisions can trigger polarization of stationary cells. Since such collisions occur more frequently at higher densities, this causes a larger fraction of cells to be motile (\myfig{liver}{E}). Thus, if Trms can help each other to move, a larger Trm population would screen the environment more efficiently than expected from the population size alone.

Notably, the mean speed of Trms peak at $\sim$10.5 μm/min (\myfig{liver}{F}), matching average speeds in our lymph node simulations -- but only at densities where the arrest coefficient is close to 0. In two-photon experiments, liver Trm speed is reported to be significantly lower than T cells in lymph nodes \cite{holz2018, hor2015spatiotemporally}; we can reproduce this in our simulations with moderate cell density, which leads to both a realistic slowdown, and higher arrest coefficient. The simulated liver Trm speed and arrest coefficient closely matches previously reported values in literature \cite{mcnamara2017up}, where they were used to quantify observed characteristic patrolling behavior of Trms.

These liver experiments demonstrated that our system allowed us to combine \ac{3D} imaging data with a relatively realistic model of polarizing cell motility. The resulting simulation produced an interesting phase transition of Trm screening efficiency.  While we do not claim that such an effect is present in the real liver -- more detailed tuning of the motility model to match the characteristics of real sinusoidal Trm cells would be out of scope for this simple proof of concept -- this does illustrate how detailed models such as the \ac{CPM} generate emergent behavior that is not an immediately obvious consequence of the model ingredients and can generate interesting biological hypotheses worthy of further exploration.

\subsection*{A T~cell collective moves robustly in a realistic, organ-scale model of the lymph node}

In \acp{LN}, millions of T~cells move in search of antigen-presenting cells \cite{Mandl18036, Textor2014}, and finding the cognate antigen is a key early step in adaptive immune responses \cite{Gasteiger2016}.  T~cell motility in the \ac{LN} is shaped by the high deformability of the cells \cite{10.3389/fimmu.2015.00586} and their interactions with a complex stromal environment, the fibroblastic reticular cell network \cite{bajenoff2006stromal, 10.1371/journal.pbio.2000827}. Given the \ac{LN}'s unique structure as an organ consisting mainly of moving cells, it presents a challenging test case for any tissue simulation system. Indeed, many models of \ac{LN} T~cell dynamics already exist, but they were often limited in either cell count and simulation volume, or in realism of cell morphology and LN environment.  For example, T~cell interactions with a synthetic stromal structure have been investigated using \ac{3D} \ac{CPM} simulations of thousands of cells \cite{beltman}, while collective motility at a more realistic scale of hundreds of thousands of cells has been investigated with non-morphological agent-based models \cite{Bogle2010, Gong2013}.  We were therefore interested in our system's ability to simulate a realistically large collective of millions of realistically shaped T~cells. As in the previous use case, we used microscopy data to initialize -- and parameterize -- our model. 

\begin{figure}[]
\centering
\includegraphics{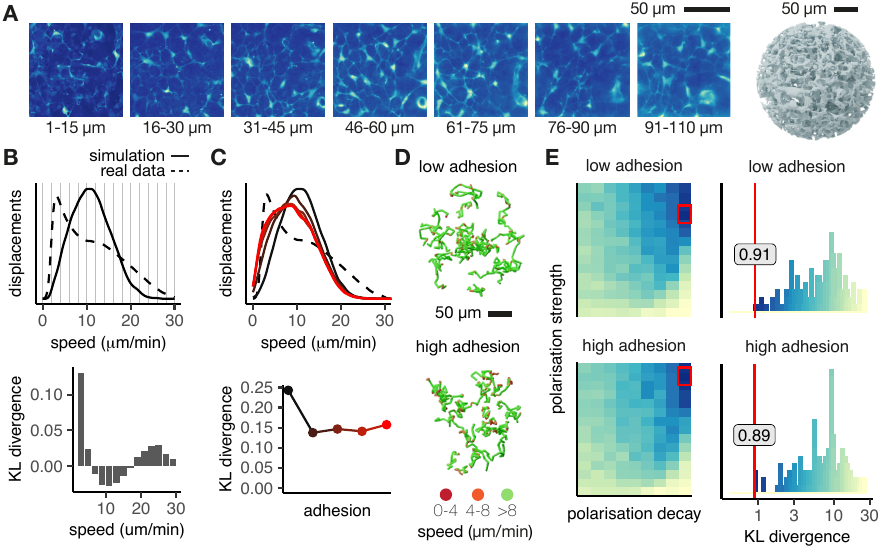}
      \caption{{\bf Reproducing \emph{in vivo} T~cell motility characteristics in a \ac{CPM}.} We compare intravital data from Mandl \emph{et al.} \cite{Mandl18036} to the motion of  50,135 \ac{CPM} cells moving in a spherical structure (256μm diameter).
      {\bfseries (A)} A 130x130x110 μm\textsuperscript{3} scan of the stromal structure (fibroblastic reticular cells) in the paracortical T~cell zone of a murine \ac{LN}. The images show average intensity projections of different depth ranges. After segmentation and mirroring, these data are used to fill a spherical simulation volume.
      {\bfseries (B)} Speeds observed in simulated data are used to construct a probability distribution (upper panel, solid), which we compare to the empirical distribution (upper panel, dashed) using  the \ac{KL} divergence metric. To this end, we bin the speed distributions, compute the divergence in each bin, and sum the values for each bin (lower panel). 
      {\bfseries (C)} Increased adhesion of T~cells to stromal cells causes the speed distributions to be more right-skewed (top) and match real data more closely (bottom). 
      {\bfseries (D)} Increased adhesion can slow T~cell paths non-uniformly by creating longer and/or more frequent pauses (red) but does not prevent rapid motion altogether. Two representative tracks are shown. 
      {\bfseries (E)} A grid search of motility parameters for cell polarization reveals which parameter ranges best fit the intravital data. Both low and high adhesion simulations have similar fit landscapes; the best fit is found in a high adhesion simulation, although the difference to the low-adhesion best fit is small (1 simulation per parameter combination, best fit: red box/red line). \ac{CPM} parameter settings can be found in Table~\ref{table_model_settings_liver_ln}.}
\label{ln_grid_search}
\end{figure}

We constructed a \ac{LN} simulation using the work of Beltman \emph{et al.} \cite{beltman} as a starting point. To improve realism, we used an imaged \ac{LN} section to initialize the stromal structure (\myfig{ln_grid_search}{A}), we used the Act-\ac{CPM} extension for cell motility \cite{actmodel}, and we fit the two Act motility parameters directly to intravital imaging data (\myfig{ln_grid_search}{B-E}). When fitting these parameters, we aimed to reproduce the intermittent stop-and-go motion \cite{miller2003autonomous} observed in real data whilst also quantitatively approximate the real speed distribution. Our system allowed fitting to be performed on realistically large simulations (256μm diameter simulation volume, containing 50,135 T~cells, 20 minutes simulated time). Crucially, we did not use periodic boundaries which, in dense simulations, can lead to artificial global synchronisation of motion in the cell collective. 

We used the \ac{KL} divergence to compare simulated to real speed distributions (\myfig{ln_grid_search}{B}). Increased adhesion of T~cells to the stromal structure made the speed distribution more right-skewed (\myfig{ln_grid_search}{C}, top), which was more consistent with real data (\myfig{ln_grid_search}{C} bottom). For most examined parameter settings,  we found the pronounced variations in cell speed along cell trajectories that are also observed in real data (\myfig{ln_grid_search}{D}). When we performed grid searches of motility parameters at two different stromal adhesion strengths (\myfig{ln_grid_search}{E}), both grid searches had broadly similar fit landscapes, and the overall lowest \ac{KL} divergence values were very similar. Qualitatively, we found more right-skewed and bimodal speed distributions in high-adhesion than low-adhesion simulations for a subset of the motility parameters (\mysupp{grid_search_distributions}). Based on these fitting results, we chose the best fit motility parameters of the high adhesion setting for our further simulations.

\begin{figure}[]
\centering
\includegraphics{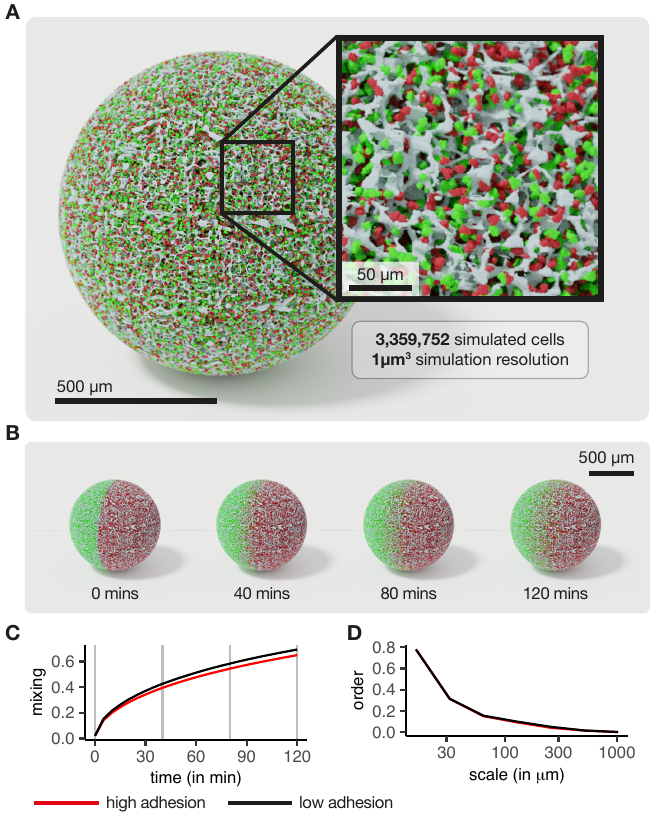}
      \caption{{\bf Mixing of motile T~cell collectives in a realistic-scale \ac{LN} simulation.} 
      {\bfseries (A)} Spherical \ac{LN} simulation at 1mm diameter containing 3,359,752 T~cells and stromal cells (grey; \myref{ln_grid_search}{A}). For clearer visualization, only 10\% of the simulated cells are shown, and red and green cells are equally parameterized although coloured differently.
      {\bfseries (B)} The simulation is initialized with two equally parameterized populations each starting at one half of the sphere. Mixing is quantified by tracking the fraction of cells having at least one neighbour of the other population.
      {\bfseries (C)} Global mixing is slower in the high-adhesion simulation (red) than at low adhesion (black), although the difference is minor. Both simulations reach 50\% mixing in \textasciitilde{}1h. 
    {\bfseries (D)} Cells do not synchronize their directionality much at longer spatial scales, and this absence of long-range synchronization is the same for low and high adhesion. \ac{CPM} parameter settings can be found in Table~\ref{table_model_settings_liver_ln}.}
\label{ln_crowd}
\end{figure}

We next investigated organ-scale collective T~cell motility by scaling up our simulated volume to 1mm diameter, now containing 3,359,752 T~cells (\myfig{ln_crowd}{A}). While this is not entirely realistic -- a real \ac{LN} has a more complex compartmental structure, while we simulate only a T~cell zone -- our main intention was to determine the spatial range of collective behavior that can now be investigated with our system. We simulated one virtual day in 1.9 days of wall clock time -- where serial implementations were projected to have taken a full 1.4 years (\myfig{cpuscaling}{A}).

We used a mixing simulation of two differentially labelled populations as an easily observable proxy for collective search (\myfig{ln_crowd}{B}). In the simulation, it takes about 1 hour to reach a state where 50\% of the cells are next to at least one cell of the other population (\myfig{ln_crowd}{C}). While stromal adhesion influences intra-cellular stop-and-go motility, it has minimal influence on collective motion: high and low stromal adhesion simulations have broadly similar mixing speeds. While collective motility is slightly ordered at cell-scale distances, no ordering persists at organ-scale (\myfig{ln_crowd}{D}).

In summary, we used an organ-scale \ac{LN} simulation as a challenging test case: an environment known for complex, large scale collective motility dynamics. Our system allowed us to integrate of microscopy data directly into our simulation, and to fit motility parameters directly on  two-photon data. We investigated the behavior of the fitted cell population at organ-scale (millions of cells). We hope that this will inspire further research on large cell collectives -- integrating various sources of empirical data, at a high level of realism, and at realistic scale.

\section*{Discussion}

We have implemented a parallel version of the \ac{CPM} that, in some experiments, was 3 to 4 orders of magnitude faster on \acp{GPU} than a state-of-the-art serial \ac{CPM}. To achieve this,  we designed novel schemes for lattice work splitting and synchronization, maintaining global state, efficient \ac{GPU} memory transfer, and multi-\ac{GPU} execution.  We found that preservation of local lattice update statistics was sufficient to maintain high-level system dynamics, despite removing the sequential nature of the original \ac{CPM} algorithm.

In our work we aimed to ensure that important properties of \ac{CPM} simulation trajectories were maintained. Some previous \ac{CPM} parallelisation efforts instead focused on equilibrium state distributions as important quantities. For example, Chen \emph{et al.}\cite{CHEN2007670} argue that their parallel \ac{CPM} maintains certain equilibrium properties. While such arguments are frequently applied in the analysis of energy-based models in physics, Voss-B\"ohme \emph{et al.} showed that these are less relevant for the \ac{CPM} because the Metropolis algorithm used in the \ac{CPM} drastically alters longtime behavior of the model \cite{criticalAnalysisCPM}.

While we have shown that the same range of behaviors is present in our parallel \ac{CPM} as in a serial \ac{CPM} in our benchmark scenarios, this would still need to be carefully checked when re-using simulation parameters that were originally fitted on a serial implementation. Specifically, researchers might want to re-fit their simulation parameters when using our system instead of a serial version -- advice that is sensible in general when moving between different implementations of the same model formalism, especially with complex models such as the \ac{CPM} where implementation details might anyway differ subtly. Furthermore, the observed relation between model kinetics and performance (\myfig{performance}{E}) also suggests it might be fruitful to explore more aggressive parallelisation settings that might yield further speed-up whilst maintaining desired simulated behavior. 

Through accelerating the \ac{CPM} on accessible \ac{GPU} hardware, we have increased the range of feasible simulation sizes -- because our scheme fully utilises all \ac{GPU} cores, simulation size is now mostly limited by the available \ac{GPU} memory. As memory per \ac{GPU} increases, our scheme will easily scale to larger simulations; our multi-\ac{GPU} prototype scheme will allow for further increase of simulation volume. Our multi-\ac{GPU} scheme could be further extended to include inter-\ac{GPU} communication over MPI \cite{MPI}, allowing our system to be used as a fundamental building block for a \ac{GPU} cluster-based version of the \ac{CPM} -- potentially scaling our 1mm\textsuperscript{3} simulations further to 1cm\textsuperscript{3} scale.

Large-scale multicellular simulations are often performed using more coarse-grained agent-based model environments \cite{Bogle2010, Gong2013, VanLiedekerke2015, reviewCellComputationalModeling}. Our high-performance \ac{CPM} now allows researchers to perform simulations that are morphologically much more detailed and allow the investigation of new questions that require taking cell shape into account, such as migration in complex environments or morphogenesis. We hope that our example simulations of liver and \ac{LN} environments will serve as inspiration for many such spatially resolved tissue and organ simulations in the future. Similar approaches to the one shown here could be useful for parallelisation of \ac{CPM} variants or other sequential lattice-based models \cite{thuroff}.

\clearpage

\section*{Tables}

\begin{table}[ht!]
\centering
\caption{
{\bf Model settings for 2D simulations.} 
For all 2D simulations we use $V_{\mathrm{target}} = 2500$, $A_{\mathrm{target}} = 600$, $\lambda_{\mathrm{perimeter}} = 2$, $J_{\mathrm{cell}, \mathrm{ECM}}=10$ and $T=10$. The vector field in the 2D circular chemotaxis simulations (\myref{fig6}{A}) consists of vectors at a 90 degree angle to radial vectors from points in the field to the simulation center, with all vectors in the field normalized to length 1. 
}
\begin{tabular}{l | l | l | l | l}
    &
    $\lambda_{\mathrm{volume}}$ & 
    $\lambda_{\mathrm{act}}$ &
    $\mathrm{Max}_{\mathrm{act}}$ &
    $\lambda_{\mathrm{chemotaxis}}$ 
    \\ \hline
    Fig~\ref{cpuscaling}B & 20, 30, 40 & 200, 250, 300, 350, 400 & 50 & --- \\
    Fig~\ref{fig6}A, Fig~\ref{performance}E & 20 & --- & --- & 10 \\
    Fig~\ref{fig6}B & 20 & 50-800, in increments of 50& 50 & ---
\end{tabular}
\label{table_model_settings_2d}
\end{table}

\begin{table}[ht!]
\centering
\caption{
{\bf Model settings for 3D simulations.} For all simulations $V_{\mathrm{target}} = 150$, $\lambda_{\mathrm{volume}} = 25$, $A_{\mathrm{target}} = 1500$, $\lambda_{\mathrm{perimeter}} = 0.2$, $J_{\mathrm{cell},\mathrm{cell}}=10$ and $T=20$. The time resolution is 1 second per Monte Carlo step.
}
\begin{tabular}{l | l | l | l | l }
    &

    $\mu$ & 
    $\Delta t$ & 
    $\lambda_{\mathrm{persistence}}$ & 

    $J_{\mathrm{cell},\mathrm{ECM}}$ 

    \\ \hline
    Fig~\ref{cpuscaling}A &  --- & --- & --- & 10 \\
    Fig~\ref{fig6}C, Fig~\ref{performance}D &  0.95 & 15 & 40 & 0 \\

\end{tabular}
\label{table_model_settings_3d}
\end{table}

\begin{table}[ht!]
\centering
\caption{
{\bf Model settings for liver and lymph node simulations.} For all simulations $V_{\mathrm{target}} = 150$, $\lambda_{\mathrm{volume}} = 25$, $A_{\mathrm{target}} = 1500$, $\lambda_{\mathrm{perimeter}} = 0.2$, $J_{\mathrm{cell},\mathrm{cell}}=10$, $J_{\mathrm{cell},\mathrm{ECM}} = 0$, and $T=20$. The time resolution is 1 second per Monte Carlo step.
}
\begin{tabular}{l | l | l | l  }
    &

    $\lambda_{\mathrm{act}}$ &
    $\mathrm{Max}_{\mathrm{act}}$ &

    $J_{\mathrm{cell},\mathrm{FRC}}$

    \\ \hline

    Fig~\ref{liver} &  2600& 10&  0\\

    Fig~\ref{ln_grid_search}B &  2600& 10 & 0\\

    Fig~\ref{ln_grid_search}C &  2600 & 10&  -400, -300, -200, -100, 0 \\

    Fig~\ref{ln_grid_search}D &  2600& 10 & -400, 10 \\

    Fig~\ref{ln_grid_search}E &  200-3000, in increments of 200 & 
		10, 15, 20, 25, 30, 40, 50, 60, 70 & -3000, 10 \\

    Fig~\ref{ln_crowd}C &  2600& 10&  -800, 0\\

    Fig~\ref{ln_crowd}D &  2600& 10&  -800 \\

\end{tabular}
\label{table_model_settings_liver_ln}
\end{table}

\clearpage

\section*{Acknowledgments}

This research was supported by Young Investigator Grant 10620 from KWF Kankerbestrijding and HFSP program grant RGP0053/2020 (to JT). JT was also supported by NWO Vidi grant VI.Vidi.192.084. We thank Franka Buytenhuijs for giving valuable feedback on this manuscript, and we thank L.W.A. Stuurman and Pedro Aroca Lara for testing early versions of this simulation system. We also gratefully acknowledge the support of NVIDIA Corporation with the donation of the Titan V GPU used for this research. The funders had no role in design, execution, and interpretation of the research reported in this manuscript.

\section*{Author contributions}

SS and JT conceived the study. SS designed the parallel CPM algorithms. SS performed the research and analysed the data. SS and JT wrote the manuscript. JT supervized the project. SM provided critical feedback and provided the liver and lymph node imaging data. SD performed image acquisition of the liver and \ac{LN} data. All authors critically revized the manuscript for intellectual content.

\section*{Declaration of interests}

The authors declare no competing interests.

\clearpage

\section*{Materials and Methods}

\subsection*{Cellular Potts model and model variants}
The CPM is a lattice based formalism for simulating biological cells. Simulations proceeds with a Markov chain Monte Carlo process, where a single position at a time can transition to one of the states of its neighbours. The transition probability is based on the change of a Hamiltonian energy function. The original CPM has
\begin{eqnarray}
\label{eq:cpm}
    \mathcal{H} = \sum_{i,j} J_{\tau(\sigma_i),\tau(\sigma_j)}(1-\delta_{\sigma_i, \sigma_j}) + \sum_{\sigma} \lambda_{\mathrm{volume}}(\tau_\sigma)(V(\sigma) - V_{\mathrm{target}} (\tau_{\sigma}))^2
\end{eqnarray}
as its Hamiltonian, with $i,j$ being possible neighbour locations in the lattice, $\sigma_i$ the cell present at location $i$, $\tau$ is the type of a particular cell, $\delta$ is the Kronecker delta, $V(\sigma)$ is the number of lattice sites a cell $\sigma$ occupies, and $\lambda_{\mathrm{volume}}(\tau)$ how strongly a volume constraint contributes to the Hamiltonian for a particular cell type. In all our simulations a Moore neighbourhood was used. Transitions stochastically succeed based on the change in energy $\Delta \mathcal{H}$, using a Boltzmann probability distribution 
\begin{eqnarray}
\label{eq:boltzmann}
    P(\Delta \mathcal{H}) = 
     \begin{cases}
         1 & \text{if $\Delta \mathcal{H} < 0$,} \\
         e^{-\Delta \mathcal{H} / T} & \text{if $\Delta \mathcal{H} \geq 0.$}
    \end{cases}       
\end{eqnarray}
By adding extra terms to the Hamiltonian, constraints can be added to cells and new types of behavior introduced to the model. When cell shape needs to be parameterized beyond just cell size, a perimeter constraint \cite{OUCHI2003451} is often introduced,
\begin{eqnarray}
\label{eq:perimeter}
    \mathcal{H}_{\mathrm{perimeter}} = \sum_{\sigma} \lambda_{\mathrm{perimeter}}(\tau_\sigma)(A(\sigma) - A_{\mathrm{target}}(\tau_{\sigma}))^2.
\end{eqnarray}
Here $A(\sigma)$ represents the number of interfaces between lattice sites of a cell with sites not part of the same cell.

We also implemented several motility models. The first allows a vector field to bias the energy change $\Delta \mathcal{H}$ during a copy attempt
\begin{eqnarray}
\label{eq:chemotaxis}
    \Delta \mathcal{H}_{\mathrm{chemotaxis}} = \lambda_{\mathrm{chemotaxis}}(\tau_\sigma)  \hat{x}_{i \rightarrow j} \cdot \vec{C_i}.
\end{eqnarray}
Here $\hat{x}_{i \rightarrow j}$ is the unit vector pointing in the direction of a copy attempt from position $i$ to $j$, and $C_i$ the value of a vector field at position $i$. In this paper, $\vec{C_i}$ is always a field of unit vectors. 

For the model where cells exhibit persistent movement based on one vector per cell, we add 
\begin{eqnarray}
\label{eq:persistence}
    \Delta \mathcal{H}_{\mathrm{persistence}} = \lambda_{\mathrm{persistence}}(\tau_\sigma)  \hat{x}_{i \rightarrow j} \cdot \hat{p}_{\sigma, t}
\end{eqnarray}
    to the Hamiltonian delta. The vector $\hat{p}_{\sigma, t}$ is the persistence direction of cell $\sigma$ at time $t$, which is updated iteratively at every \ac{MCS} with
\begin{eqnarray}
\label{eq:persistence_update}
    \vec{p}_{\sigma, t} = (1-\mu_\tau) 
    \frac{\vec{X}_{\sigma,t} - \vec{X}_{\sigma,t-\Delta t}}
    {\lVert \vec{X}_{\sigma,t} - \vec{X}_{\sigma,t-\Delta t} \rVert}
    + \mu_\tau \hat{p}_{\sigma, t-1}, \\
    \hat{p}_{\sigma, t} =  
    \frac{\vec{p}_{\sigma, t} }
    {\lVert \vec{p}_{\sigma, t} \rVert}.
\end{eqnarray}
Here $\mu_\tau$ is the strength of directional persistence for cell type $\tau$, $\vec{X}_{\sigma, t}$ is the centroid location of cell $\sigma$ at time $t$ and $\Delta t$ is the number of MCSs we look back for to determine the current direction of a cell. 

Finally, we implement the Act model \cite{actmodel} that allows for localized polarization. For each position $i$ we track an activity value each time said location is updated, here labelled as $\mathrm{Act}(i)$. This localized `memory' is initialized at $\mathrm{Max}_{\mathrm{act}}$, and decays by 1 each timestep until it reaches 0. The Hamiltonian delta is adapted by 
\begin{eqnarray}
\label{eq:act}
    \Delta \mathcal{H}_{\mathrm{act}} = 
    \frac{\lambda_{\mathrm{act}}(\tau_\sigma)} 
    {\mathrm{Max}_{\mathrm{act}}(\tau_\sigma) } (
    \mathrm{GM}_{\mathrm{act}}(i) -
    \mathrm{GM}_{\mathrm{act}}(j)
    ),
\end{eqnarray}
making copies that originate from areas of a cell with a high activity value more likely. The activity is calculated for a full Moore neighbourhood using a geometric mean
\begin{eqnarray}
\label{eq:mean}
    \mathrm{GM}_{\mathrm{act}}(i) = 
    (\prod_{n \in N(i)} \mathrm{Act}(n)
    )^{1/\lvert N(i) \rvert}.
\end{eqnarray}
$N(i)$ are all neighbours of pixel or voxel $i$.

\subsection*{Optimized \ac{CPU} implementation of \ac{CPM}}

Besides a high-performance GPU implementation, we also developed our own \ac{CPU} implementation. This served both as a reference implementation for model behavior and internal implementation details for us, and as a reference performance benchmark. We strived to reduce the major performance bottlenecks we could see in profiling. The implementation tracks lattice sites on cell boundaries, and only tries copy attempts involving such positions; a common optimisation, which is also implemented in Morpheus. The tracking implementation uses a high-performance hashmap implementation \cite{bytetell}. We use a custom high-performance pseudorandom number generator, xoshiro256 \cite{Blackman2021}. The lattice data is stored with a Z-order space filling curve, allowing data that is close in the simulation space also to be close in the linear memory space, maximizing cache efficiency. Memory addresses are calculated using intrinsic bit interleaving instructions from the x86 bit manipulation instruction set extension. 

\subsection*{\ac{GPU} implementation details}

Our CPM was implemented using CUDA; concepts equivalent to ones described here exist in non-CUDA GPU hardware. A GPU consists of multiple streaming multiprocessors (SMs). Each SM runs multiple threads in parallel; contrasting with CPU threads, threads on the same SM execute in lockstep, in a Single Instruction Multiple Threads approach. A single \ac{GPU} process -- called a kernel -- can consist of up to millions of threads.

GPUs have relatively slow global memory available to all threads, and much faster shared memory local to SMs. Global memory transfer speed is often the main performance bottleneck for GPU programs. Threads are bundled in thread blocks (TBs); each TB is bound to a specific SM. By caching data on their SM in shared memory, TBs can greatly reduce global memory transfer bottlenecks. 

Memory transfer bottlenecks can be further reduced by two strategies: read coalescing and high SM occupancy. Read coalescing has sequential threads in a TB do contiguous reads of global memory; contiguous reads are needed to fully utilise the memory bus bandwidth. High occupancy is achieved when a high number of threads are scheduled to execute per SM; the SM scheduler will switch out threads when they are blocked on memory transfers, ensuring a high number of in-flight memory instructions at a time. 

We propose a scheme of memory transfer, local SM execution and synchronization (\myfig{parallel_scheme}{D}). We focus on caching just the grid of cell ids, as this data is always needed, and can be used to quickly reject copy attempts internal to a cell. The simulation grid is divided in subsections, with regions larger than earlier checkerboard subdivisions; each subsection is assigned to a TB, subsection size is chosen so active regions fit in shared memory. Each TB determines which checkerboard sections are active in its region; active sections are cached in shared memory using coalesced reads. 

After caching -- intended to speed up reads -- the simulation is run, and successful copy attempts write new values to both shared and global memory. Alternatively, changes could be cached in shared memory and written back coalesced to global memory; in our testing, an immediate shared/global write yielded better performance. If a TB is done with simulation execution, it immediately starts a memory request for the next active region, ensuring maximum occupancy. All threads in the kernel are synchronized after new memory transfers are finished, ensuring maximum overlap of code execution and memory transfer.

\ac{CPM} literature offers a wide range of model variants and extensions, of which we implement a modest subset. This comes with potential performance issues: even when extensions are disabled through branching statements, on \acp{GPU} such branching code still incurs a significant performance loss compared to CPUs. We therefore pre-generated specialized versions of the kernel for a range of different extensions. 

We use a custom high-performance random number generator on the GPU \cite{10.1145/1570256.1570353}.

\subsection*{Composite variable tracking}
\label{composite_variable}
Several CPM extensions introduce additional global variables. Some of these variable need to be kept in-sync with each other -- one variable being slightly out of date with others is not tolerated. One example is tracking of a center-of-mass centroid; for 2D this can be tracked as a sum of x and y values, and corresponding count of lattice locations occupied (\myfig{parallel_scheme}{C}). These are integer values: tracking centroid locations as floating-point would cause values to drift due to rounding errors. The centroid value is calculated by 
\begin{eqnarray}
\label{eq:centroid}
    \{x, y\} = 
    \frac{1}{\lvert \bullet \rvert} \times 
    \{
        x_{\mathrm{sum}}, y_{\mathrm{sum}}
        \}.
\end{eqnarray}
Online updates change both a cell's centroid position and size. When a new lattice position $\{x_{\mathrm{position}}, y_{\mathrm{position}}\}$ is added to a cell, its centroid's $x$ value can be updated with
\begin{eqnarray}
\label{eq:centroid_update}
    x_{\mathrm{new}} 
    = 
    \frac{x_{\mathrm{old}} \times \lvert \bullet \rvert_{\mathrm{old}} + x_{\mathrm{position}}}
    {\lvert \bullet \rvert_{\mathrm{new}}}.
\end{eqnarray}

Cell sizes are stored interleaved with sum values; sizes and sums are both stored as 32-bit unsigned integers. The interleaved storage enables use of 64-bit operations to update both: a copy-and-swap (CAS) operation is used, which atomically updates sum and cell size values simultaneously. This scheme also handles periodic boundaries, by adding or subtracting $\mathrm{dimension} \times \lvert \bullet \rvert$ to the sum values before executing the CAS as needed, to keep centroids within simulation bounds.

\subsection*{Multi-\ac{GPU} execution}

We devized a scheme to execute our \ac{CPM} on multiple \acp{GPU} (\myfig{parallel_scheme}{E}); this scheme is intended as a prototype, and as such not as optimized as other parts of our \ac{CPM}. Simulation space is cut into equal regions, each region is assigned to a \ac{GPU}. A region's lattice data, such as the cell ids, are sent to each \ac{GPU}, with an additional border layer. Global variables such as cell sizes are tracked on each \ac{GPU} as deltas of the value since last inter-\ac{GPU} data synchronisation. Every four \acp{MCS}, these deltas are gathered from all \acp{GPU}, deltas are integrated, and updated values distributed to each \ac{GPU}. Multi-\ac{GPU} speedup is mostly hampered by frequent or slow inter-\ac{GPU} communication; inter-\ac{GPU} synchronization after every \ac{MCS} negates most speedup, and a faster \ac{GPU}-to-\ac{GPU} interlink would likely increase performance. 

\subsection*{Hardware}
The \ac{CPU}-based simulations, using our custom optimized \ac{CPU} implementation and Morpheus 2.2.6 (the last non-parallelized version of Morpheus), and most \ac{GPU} simulations were run on a system with a Nvidia Titan V, an AMD 2950x \ac{CPU} and 128 GB of RAM at 2133 MHz. The only exceptions to this are the multi-\ac{GPU} simulations (\myfig{parallel_scheme}{E}), which were run on a system with four Nvidia Geforce RTX 2080 Ti GPUs, an Intel i9-9820x \ac{CPU} at 3.3 GHz, and 128 GB of RAM at 2133 MHz. 

\subsection*{Imaging of thick lymph node sections}
Na\"ive inguinal \acp{LN} were harvested from CCL19-EYFP mice \cite{chai} and fixed with 4\% paraformaldehyde (PFA, Electron Microscopy Sciences) for 1 hour at 4\textdegree C. Following a brief wash with PBS, \acp{LN} were embedded in 2\% agarose and 250-300 µm tissue sections were generated using a vibratome (Leica). \ac{LN} sections were subsequently immersed in a Special Wash Buffer (Baxter water, 0.2\% BSA, 10\% 10X PBS, 0.2\% Tween-20, 0.2\% Triton X-100, 0.2\% 10\% SDS) at 4\textdegree C overnight, followed by staining with a rabbit anti-GFP Ab (Thermo Scientific, 1:100, diluted in blocking buffer made with Baxter water, 0.2\% BSA, 10\% 10X PBS, 0.1\% Tween-20, 0.2\% 10\% SDS) at 4\textdegree C for three days. After incubation, \ac{LN} sections were thoroughly washed in PBS and mounted in 85\% glycerol in PBS with glass coverslips. Thick \ac{LN} sections were imaged using a LSM880 AiryscanFast confocal microscope (Carl Zeiss) with a x63/1.4NA oil immersion lens and images processed with ZEN 2010 software (Carl Zeiss).

\subsection*{Imaging of liver sinusoids}
For whole-liver imaging, AlexaFluor 647 conjugated anti-CD31 rat monoclonal antibody (Biolegend) was intravenously injected into wild-type C57BL/6 mice at 5ug for 15 minutes prior to harvesting the liver. This was followed by overnight 4\% PFA fixation at 4°C. Livers were then thoroughly washed with PBS and small cubes of liver were prepared before immersion in FUnGI \cite{Rios2019} for two days before mounting in 85\% glycerol in PBS. Thick liver sections were imaged using FVMPE-RS multiphoton microscope (Olympus) with a x20/1.05 NA water immersion lens.

\clearpage

\section*{Data availability}

The liver imaging data and the segmented version are available on Zenodo \cite{ZenodoLiver}. 
The lymph node imaging data and the segmented version are available on 
Zenodo \cite{ZenodoFRC}.
Source data and code for the figures will be published together with the final, archival version
of this manuscript. In the meantime, these data and code are available upon request to the corresponding
author and are made available to reviewers.

\section*{Code availability}

\ac{GPU} \ac{CPM} simulation code is available on GitHub at \url{https://github.com/shabaz/gpu-cpm} 
and is made available under the GNU General Public License, version 3. 
\ac{CPU} \ac{CPM} simulation code is available on GitHub at  \url{https://github.com/shabaz/cpm} 
and is made available under the GNU General Public License, version 3. 

\clearpage

\bibliographystyle{unsrtm}
\bibliography{ms}

\clearpage

\section*{List of abbreviations}

\begin{acronym}
  \acro{2D}{two-dimensional}
  \acro{3D}{three-dimensional}
  \acro{CPM}{Cellular Potts Model}
  \acro{GPU}{Graphical Processing Unit}
  \acro{CPU}{Central Processing Unit}
  \acro{MCS}{Monte Carlo Step}
  \acro{LN}{Lymph Node}
  \acro{KL}{Kullback-Leibler}
\end{acronym}

\clearpage

\section*{Supplementary Figures}

\newcounter{sfigure}
\newcounter{stable}

\renewcommand{\thefigure}{S\arabic{sfigure}}
\renewcommand{\thetable}{S\arabic{stable}}
\renewcommand{\figurename}{Supplementary Figure}
\renewcommand{\tablename}{Supplementary Table}

\stepcounter{sfigure}

\begin{figure}[ht!]
\centering
\includegraphics{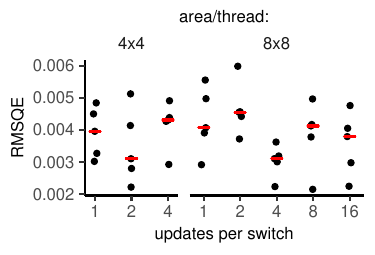}
    \caption{
    {\bf Checkerboard parallelisation scheme does not have major influence on sorting model accuracy.} We tested sorting simulations with atomic updates, using different options of parallelisation schemes, varying space assigned per thread and switching frequency of active checkerboard areas. Accuracy is measured as the root mean square error of the GPU simulation runs with 5 CPU reference runs and 5 GPU runs per parallelisation setting (median measurement marked with line), measuring homogeneous/heterogeneous cell border interfaces over time at t=0, 1, 10, 100, 1000, 10000, and 100000 seen in \myfig{sorting}. Absolute error values remain low, and there are no pronounced differences between errors for the different GPU settings (two-way ANOVA, p=0.43).}
\label{sorting_accuracy}
\end{figure}

\clearpage

\stepcounter{sfigure}
\begin{figure}[]
\centering
\includegraphics{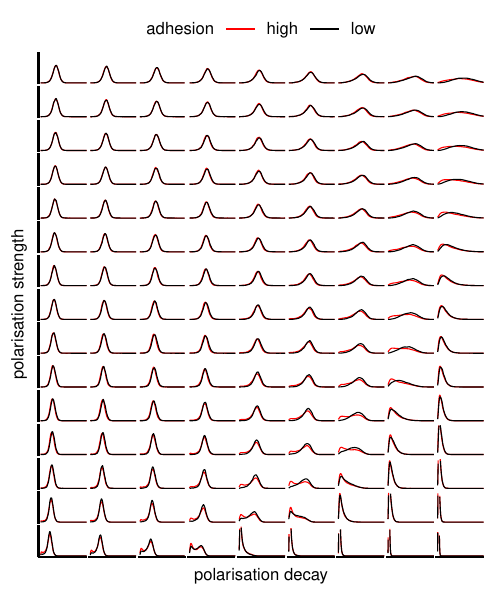}
    \caption{
  {\bf Cell speed distributions obtained when fitting \ac{LN} motility data.} 
Each plot corresponds to one cell in the grid search shown in \myref{ln_grid_search}{E}.
Polarization decay refers to the parameter $\mathrm{Max}_{\mathrm{act}}$; polarization
strength refers to the parameter $\lambda_{\mathrm{act}}$.
}
\label{grid_search_distributions}
\end{figure}

\end{document}